\documentclass[aps,pra,reprint,twocolumn,superscriptaddress,floatfix,nofootinbib]{revtex4-1}

\usepackage{amsmath}
\usepackage{amssymb}
\usepackage[utf8]{inputenc}
\usepackage[T1]{fontenc}
\usepackage[separate-uncertainty=true]{siunitx}
\usepackage[export]{adjustbox}
\usepackage{float}
\usepackage[resetlabels]{multibib}
\newcites{SI}{SI}

\begin{document}

\author{Elia Strambini}
\email{elia.strambini@sns.it}
\affiliation{NEST, Istituto Nanoscienze-CNR and Scuola Normale Superiore, I-56127 Pisa, Italy}

\author{Andrea Iorio}
\email{andrea.iorio@sns.it}
\affiliation{NEST, Istituto Nanoscienze-CNR and Scuola Normale Superiore, I-56127 Pisa, Italy}

\author{Ofelia Durante}
\affiliation{Dipartimento di Fisica ``E. R. Caianiello'', Università di Salerno, 84084 Fisciano (Salerno), Italy}

\author{Roberta Citro}
\affiliation{Dipartimento di Fisica ``E. R. Caianiello'', Università di Salerno, 84084 Fisciano (Salerno), Italy}

\author{Cristina Sanz-Fernández}
\affiliation{Centro de Fısica de Materiales (CFM-MPC), Centro Mixto CSIC-UPV/EHU, Manuel de Lardizabal 4, E-20018 San Sebastian, Spain}

\author{Claudio Guarcello}
\affiliation{Dipartimento di Fisica ``E. R. Caianiello'', Università di Salerno, 84084 Fisciano (Salerno), Italy}
\affiliation{Centro de Fısica de Materiales (CFM-MPC), Centro Mixto CSIC-UPV/EHU, Manuel de Lardizabal 4, E-20018 San Sebastian, Spain}

\author{Ilya V. Tokatly}
\affiliation{Nano-Bio Spectroscopy group, Departamento de Fısica de Materiales, Universidad del Paıs Vasco Av. Tolosa 72, E-20018 San Sebastian, Spain}
\altaffiliation{IKERBASQUE, Basque Foundation for Science - E-48011 Bilbao, Spain}
\altaffiliation{Donostia International Physics Center (DIPC), Manuel de Lardizabal 4, E-20018 San Sebastian, Spain}

\author{Alessandro Braggio}
\affiliation{NEST, Istituto Nanoscienze-CNR and Scuola Normale Superiore, I-56127 Pisa, Italy}

\author{Mirko Rocci}
\affiliation{NEST, Istituto Nanoscienze-CNR and Scuola Normale Superiore, I-56127 Pisa, Italy}
\altaffiliation{Current address: Francis Bitter Magnet Laboratory and Plasma Science and Fusion Center, Massachusetts Institute of Technology, Cambridge, MA 02139, USA}

\author{Nadia Ligato}
\affiliation{NEST, Istituto Nanoscienze-CNR and Scuola Normale Superiore, I-56127 Pisa, Italy}

\author{Valentina Zannier}
\affiliation{NEST, Istituto Nanoscienze-CNR and Scuola Normale Superiore, I-56127 Pisa, Italy}
\author{Lucia Sorba}
\affiliation{NEST, Istituto Nanoscienze-CNR and Scuola Normale Superiore, I-56127 Pisa, Italy}

\author{F. Sebastian Bergeret}
\email{fs.bergeret@csic.es}
\affiliation{Donostia International Physics Center (DIPC), Manuel de Lardizabal 4, E-20018 San Sebastian, Spain}
\altaffiliation{Centro de Fısica de Materiales (CFM-MPC), Centro Mixto CSIC-UPV/EHU, Manuel de Lardizabal 5, E-20018 San Sebastian, Spain}

\author{Francesco Giazotto}
\email{francesco.giazotto@sns.it}
\affiliation{NEST, Istituto Nanoscienze-CNR and Scuola Normale Superiore, I-56127 Pisa, Italy}
\email{f.giazotto@sns.it}

\title{A Josephson phase battery}
\maketitle

\textbf{A battery is a classical apparatus which converts a chemical reaction into a persistent voltage bias able to power electronic circuits. 
Similarly, a phase battery is a quantum equipment which provides a persistent phase bias to the wave function of a quantum circuit. It represents a key element for quantum technologies based on quantum coherence.
Unlike the voltage batteries, a phase battery has not been implemented so far, mainly because of the natural rigidity of the quantum phase that, in typical quantum circuits, is imposed by the parity and time-reversal symmetry constrains. Here we report on the first experimental realization of a phase battery in a hybrid superconducting circuit. 
It consists of an n-doped InAs nanowire with unpaired-spin surface states and proximitized by Al superconducting leads.
We find that the ferromagnetic polarization of the unpaired-spin states is efficiently converted into a persistent phase bias $\varphi_0$ across the wire, leading to the anomalous Josephson effect~\cite{buzdin_direct_2008,bergeret_theory_2015}. By applying an external in-plane magnetic field a continuous tuning of $\varphi_0$ is achieved. This allows the charging and discharging of the quantum phase battery and reveals the symmetries of the anomalous Josephson effect predicted by our theoretical model. 
Our results demonstrate how the combined action of spin-orbit coupling and exchange interaction breaks the phase rigidity of the system inducing a strong coupling between charge, spin and superconducting phase. 
This interplay opens avenues for topological quantum technologies~\cite{alicea_exotic_2013}, superconducting circuitry~\cite{linder_superconducting_2015,fornieri_towards_2017} and advanced schemes of circuit quantum electrodynamics~\cite{wallraff_strong_2004,chiorescu_coherent_2004}.}

At the base of phase-coherent superconducting circuits is the Josephson effect~\cite{josephson_possible_1962}: a quantum phenomenon describing the flow of a dissipationless current in weak-links between two superconductors. The Josephson current $I_J$ is then intimately connected to the macroscopic phase difference $\varphi$ between the two superconductors via the so called current-phase relationship (CPR) $I_J(\varphi)$. If either time-reversal ($t \rightarrow -t$) or inversion ($\Vec{r} \rightarrow -\Vec{r}$) symmetries are preserved, $I_J(\varphi)$ is an odd function of $\varphi$ and the CPR, in its simplest form, reads $I_J(\varphi)=I_C \sin(\varphi)$~\cite{golubov_current-phase_2004}, with $I_C$ being the junction critical current. This means that, as long as one of these symmetries is preserved, an open Josephson junction (JJ) ($I_J=0$) cannot provide a phase bias or, accordingly, a JJ closed on a superconducting circuit ($\varphi=0$) cannot generate current. As a consequence, the implementation of a phase battery~\cite{pal_quantized_2019} is prevented by these symmetry constraints which impose a rigidity on the superconducting phase, a universal constraint valid for any quantum phase~\cite{yacoby_phase_1996,strambini_impact_2009} .

\begin{figure*}[ht]
\includegraphics[center]{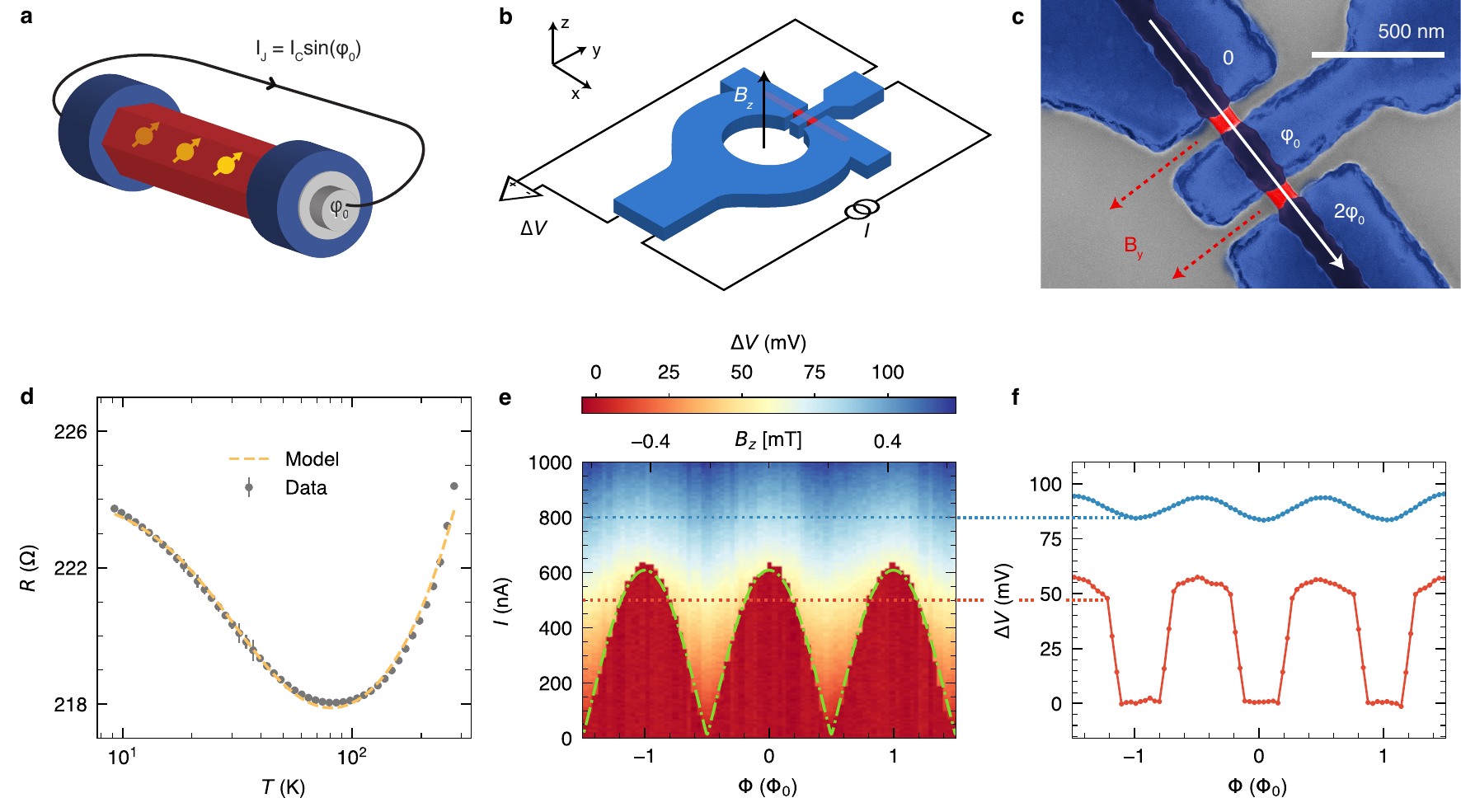}
\caption{\textbf{Josephson phase-battery device.} 
\textbf{a,} Conceptual scheme of a Josephson phase battery composed by an InAs nanowire (red) embedded between two superconducting poles (blue) converting the spin polarization of surface unpaired spins (yellow) into a phase bias $\varphi_0$. The battery, inserted into a superconducting circuit, can generate a supercurrent proportional to $\sin{(\varphi_0})$.
\textbf{b,} Schematic illustration of the hybrid InAs nanowire-aluminum SQUID interferometer used to quantify the phase bias $\varphi_0$ provided by the two JJs (in red). The ring lies in the $x-y$ plane with the nanowire parallel to the $x$-axis. A magnetic field $B_z$ piercing the ring is used to modulate the SQUID critical current ($I_S$) measured with a 4-wire setup. 
$I$ is the current flowing through the interferometer, whereas $\Delta V$ is the resulting voltage drop across the device.
\textbf{c,} False-colored SEM image of the active region of the phase battery composed by the two $\varphi_0$-junctions. $B_y$ is the in-plane magnetic field orthogonal to the nanowire. 
\textbf{d,} Temperature dependence $R(T)$ of the normal-state resistance of the interferometer showing a Kondo upturn at low temperatures which is consistent with a background of magnetic impurities. From the fit (yellow curve) we estimate a spin density of $\sim 4$~ppm. Error bars indicate the resistance standard deviations between two temperature points. \textbf{e,} Voltage drop $\Delta V$ measured across the SQUID versus current bias $I$ and magnetic flux (field) $\Phi$($B_z$). The periodicity of $I_S$ as a function of $B_z$ translates into a periodicity on the flux $\Phi = B_z A$ through the ring with an effective area $A\simeq\SI{4.9}{\square\micro\meter}$, consistent with the area estimated from Fig.~S9 in the Supplementary Information. The green line is the best-fit of the SQUID critical current $I_S(\Phi)$ (Eq.~\eqref{eq_squid}) defined by the interface between the dissipationless (red area) and the dissipative regime (coloured).
\textbf{f,} Traces of $\Delta V(\Phi)$ from \textbf{e,} measured for two selected values of the current bias (below and above 2$I_C$) which demonstrate the $\Phi_0$-periodicity both in the dissipationless and in the dissipative regime. Data in \textbf{e} and \textbf{f} were recorded at a bath temperature of $\SI{50}{\milli \kelvin}$.}
\label{fig1}
\end{figure*}

The break of time-reversal symmetry (alone) maintains the phase-rigidity but enables two possible phase shifts $0$ or $\pi$ in the CPR. The $0$-$\pi$ transition has been extensively studied in superconductor/ferromagnet/superconductor junctions~\cite{ryazanov_coupling_2001,golubov_current-phase_2004} which has applications in cryogenic memories~\cite{baek_hybrid_2014,gingrich_controllable_2016}. On the other hand, if both, time-reversal and inversion symmetries are broken a finite phase shift $0<\varphi_0<\pi$ can be induced~\cite{bergeret_theory_2015,silaev2017anomalous} and the CPR reads:
\begin{equation}
I_J(\varphi)=I_C\sin{(\varphi +\varphi_0)}.
\label{eq_CPR}
\end{equation}
A junction with such CPR, defined as a $\varphi_0$-junction~\cite{buzdin_direct_2008}, will generate a constant phase bias $\varphi = - \varphi_0 $ in an open circuit configuration, while inserted into a closed superconducting loop will induce a current $I =I_C\sin{(\varphi_0)}$, usually denoted as \emph{anomalous} Josephson current. Recently, anomalous Josephson currents have been the subject of theoretical~\cite{yokoyama_magnetic_2014,pal_quantized_2019} and experimental works~\cite{szombati_josephson_2016,assouline_spin-orbit_2019,mayer_gate_2020} envisioning direct applications on superconducting electronics and spintronics~\cite{linder_superconducting_2015,pal_quantized_2019}. 

Lateral hybrid junctions made of materials with a strong spin-orbit interaction~\cite{szombati_josephson_2016,mayer_gate_2020} or topological insulators~\cite{assouline_spin-orbit_2019} are ideal candidates to engineer Josephson $\varphi_0$-junctions. The lateral arrangement breaks the inversion symmetry and provides a natural polar axis $\mathbf{{\hat z}}$ perpendicular to the current direction. Moreover, the electron spin polarization induced by either a Zeeman field or the exchange interaction with ordered magnetic impurities, breaks the time-reversal symmetry. In this case, the anomalous $\varphi_0$-shifts is ruled by the Lifshitz-type invariant in the free energy ($F_L$), which has the form~\cite{bergeret_theory_2015}: 

\begin{equation}
F_L \sim f(\alpha,h)(\mathbf{n_h}\times \mathbf{\hat z}) \cdot \mathbf{v_s},
\label{eq_FL}
\end{equation}
where $f(\alpha,h)$ is an odd function of the strength of the Rashba coefficient $\alpha$ and the exchange or Zeeman field $h$, $\mathbf{n_h}$ is a unit vector pointing in the direction of the latter, and $\mathbf{v_s}$ is the superfluid velocity of the Cooper pairs flowing in the JJ. The scalar triple product then defines the vectorial symmetries of $\varphi_0$, while the amplitude of the shift depends on sample-specific microscopic details as well as macroscopic quantities like temperature.

Driven by the geometric condition for a finite $\varphi_0$-shift (Eq.~\eqref{eq_FL}), we realized a phase battery (Fig.~\ref{fig1}a and b) consisting of a JJ made of an InAs nanowire (in red) embedded between two Al superconducting poles (in blue). The supercurrent, and hence $\mathbf{v_s}$, flows along the wire ($x$-direction) which is orthogonal to the effective $SU(2)$ Rashba magnetic field vector pointing out of the substrate plane ($z$-direction) hosting the InAs nanowire. In the same nanowire, surface oxides or defects generate unpaired spins behaving like ferromagnetic impurities (represented by yellow arrows in Fig.~\ref{fig1}a) that can be polarized along the $y$-direction to provide a persistent exchange interaction $h$ in this direction. This leads to a finite triple product in Eq.~\eqref{eq_FL} and, consequently, to the anomalous $\varphi_0$ phase bias.

An Al-based superconducting quantum interference device (SQUID) is used as a phase-sensitive interferometer made with two $\varphi_0$-JJs (in red), as shown in Fig.~\ref{fig1}b, c (see Section I of the Supplementary Information for fabrication detail). The device geometry has been conceived to maximize the symmetry of the two JJs~\cite{giazotto_josephson_2011} to accumulate the two anomalous $\varphi_0$-shifts when applying a uniform in-plane magnetic field. The anomalous phase shift in the SQUID critical current, is then given by:
\begin{equation}
I_{S}(\Phi)=2I_{C} \left|\cos{\left(\pi\frac{\Phi}{\Phi_0}+\frac{\varphi_{tot}}{2}\right)}\right|,
\label{eq_squid}
\end{equation}
where $I_{C}$ is the critical current of each JJ, $\Phi$ is the magnetic flux piercing the ring, $\varphi_{tot} = 2\varphi_{0}$ is the total anomalous phase shift in the SQUID interference pattern resulting from the $\varphi_0$-shifts in each JJs (see Section IV of the Supplementary Information for details), and $\Phi_0 = \SI{2.067e-15}{\weber}$ is the flux quantum. This model provides a good description of the SQUID interference pattern displayed in Fig.~\ref{fig1}e, which shows the voltage drop across the SQUID as a function of the out-of-plane magnetic field $B_z$ and bias current $I$. The red-colored region of Fig.~\ref{fig1}e, corresponding to zero-voltage drop, indicates the dissipationless superconducting regime and the edge of this region provides the $I_S(\Phi)$ dependence. The green line on top of the color plot is the best-fit of $I_S(\Phi)$ from Eq.~\eqref{eq_squid}, with $I_{c}\simeq \SI{300}{\nano\ampere}$ and no phase-shift $\varphi_{tot} \simeq 0$. The latter condition is consistent with the absence of the anomalous phase when the magnetic field has only a component in $\mathbf{\hat z}$ direction and the magnetic impurities are not polarized (i.e. $\mathbf{n_h}\parallel \mathbf{\hat z}$ in Eq.~\eqref{eq_FL}). Notably, there is a replica of the $I_S(\Phi)$ oscillations in the voltage drop $\Delta V(\Phi)$ when $I>I_S$, and the SQUID operates in the dissipative regime (blue region and curve in Figs.~\ref{fig1}e-f), as conventionally realized with strongly overdamped JJs~\cite{clarke_squid_2004}. This oscillation provides a complementary and fast method to quantify the SQUID phase shifts and is used in the following analysis. Additional measurements on similar devices can be found in Section VII of the Supplementary Information.

The temperature dependence of the device normal-state resistance shows an upturn below $\sim \SI{80}{\kelvin}$ (see Fig.~\ref{fig1}d) which is a clear signature of the presence of magnetic impurities that increase, at low temperature, the electron scattering events. The upturn can be well fitted by the Kondo model (yellow line of Fig.~\ref{fig1}d) for spin 1/2 of magnetic impurities with a density of $\sim 4$~ppm (see Section II of the Supplementary Information for more details on the fitting procedure). The presence of these unpaired spins can be ascribed to the nanowire surface oxides, as already observed in undoped metal oxide nanostructures~\cite{sapkota_observations_2018}, even if defects in the nanowire crystalline structure~\cite{dietl_engineering_2006} cannot be excluded a priori. Although, the amount of intrinsic magnetic impurities is not fully controllable, their presence is crucial for the operation and implementation of the phase battery, as discussed below.

\begin{figure*}[t]
\includegraphics[center]{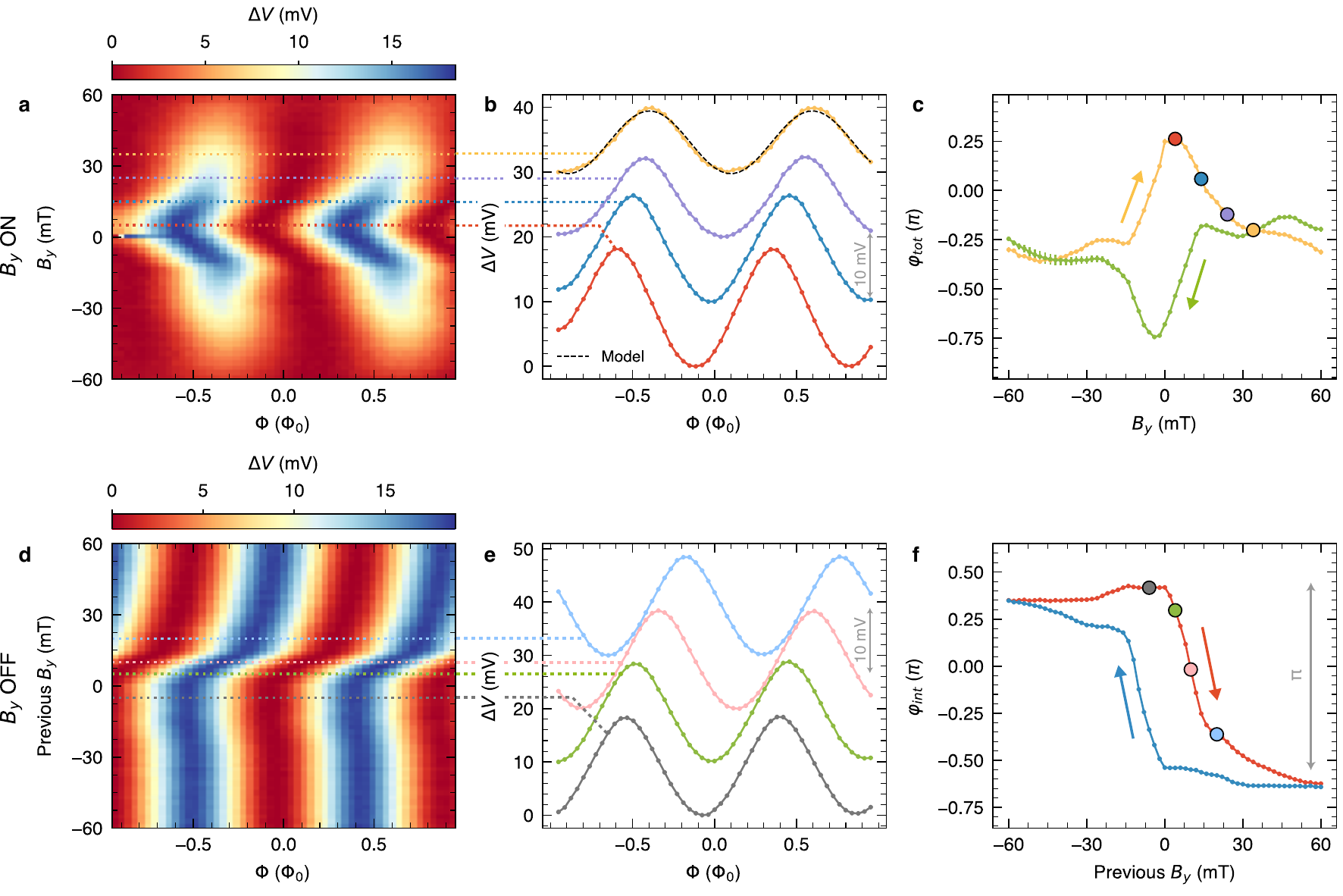}
\caption{\textbf{Charging loops of the Josephson phase battery.} 
\textbf{a,} Voltage drop $\Delta V(\Phi)$ at constant current bias $I=\SI{1}{\micro\ampere}$ versus in-plane magnetic field $B_{y}$ applied orthogonal to the nanowire axis. At large $|B_{y}|$, the amplitude of $\Delta V(\Phi)$ is lowered due to the suppression of superconductivity inside the wire. Each trace is vertically offset for clarity.
\textbf{b,} Selected traces $\Delta V(\Phi)$ extracted from \textbf{a} for different $B_{y}$. Data are vertically offset for clarity. 
\textbf{c,} Extracted phase shift $\varphi_{tot}$ from the curves in \textbf{a} with back (green) and forth (yellow) sweeps in $B_y$. The shifts are extracted by comparing the single traces to the RSJ relation $\Delta V = \tfrac{R}{2}\sqrt{I^2
-4I_C^2 \cos{(\pi \Phi / \Phi_0 + \varphi_{tot}/2)}^2}$~\cite{clarke_squid_2004}, see the dashed line in \textbf{b} which serves as an example. 
\textbf{d,} Color plot of the persistent voltage drop $\Delta V(\Phi)$ measured at $B_{y}=0$ after the magnetic field was swept to the values shown on the $y$-axis. 
\textbf{e,} Selected traces $\Delta V(\Phi)$ extracted from \textbf{d}.
\textbf{f,} Intrinsic phase shift $\varphi_{int}$ extracted from \textbf{d}. $\varphi_{int}$ stems from the ferromagnetic polarization of the unpaired spins. Error bars in c and f indicate the 1$\sigma$ standard errors resulting from the fit of the curves in b and e. All data were recorded at $\SI{50}{\milli \kelvin}$ of bath temperature.}
\label{fig2}
\end{figure*}

Following the condition imposed by a finite Liftshitz invariant term (Eq.~\eqref{eq_FL}) we apply an in-plane magnetic field orthogonal to the nanowire axis ($B_{y}$) to maximize the effect. The $I_S(\Phi)$ dependence then evolves with a clear generation of an anomalous phase shift, as presented in the panels of Fig.~\ref{fig2}. The evolution of $\Delta V(\Phi)$ as a function of $B_{y}$ ranging from $\SI{-60}{\milli\tesla}$ up to $\SI{60}{\milli\tesla}$ is visible in Fig.~\ref{fig2}a and in the selected single traces of Fig.~\ref{fig2}b. The resulting phase-shift, $\varphi_{tot}$ exhibits a non-monotonic evolution as a function of $B_y$, with a maximum shift at $B_{y}\simeq \SI{5}{\milli\tesla}$ and a saturation for $| B_{y} | \gtrsim \SI{30}{\milli\tesla}$ (yellow curve in Fig.~\ref{fig2}c). When the field is reversed, a hysteretic behavior is observed (green curve in Fig.~\ref{fig2}c), and the evolution of $\varphi_{tot}$ reverses with a minimum shift at $B_{y}\simeq \SI{-5}{\milli\tesla}$. The change of sign of the phase-shift agrees with the theoretical prediction of Eq.~\eqref{eq_FL} when $\mathbf{h}\rightarrow-\mathbf{h}$, whereas the observed hysteretic behavior suggests a ferromagnetic coupling between the magnetic impurities in the nanowire. Trivial hysteretic phase shifts induced by a trapped flux in the superconductor~\cite{PhysRevLett.104.227003} or in the SQUID ring  can be excluded (see Section VI of the Supplementary Information for more details). At low temperatures, the coexistence of Kondo and ferromagnetism is not unusual~\cite{sapkota_observations_2018} and well describes the hysteretic non-monotonic behavior observed in $\varphi_{tot}(B_y)$. Indeed, due to the antiferromagnetic nature of the Kondo interaction, the effective exchange field created by these unpaired spins is opposite to the Zeeman field generated by $B_y$ so that the two contributions are competing in the anomalous phase with a partial cancellation. 

This additional component is confirmed by the observation of an \emph{intrinsic} phase-shift, $\varphi_{int}$, which is present even in the absence of the in-plane magnetic field ($B_{y}=0$) if a finite $B_y$ has been previously applied, as shown in Figs.~\ref{fig2}d and e. Since it stems from a ferromagnetic ordering, $\varphi_{int}$ depends only on the history of $B_y$, and again, the evolution of $\varphi_{int}(B_y)$ can be extracted and is presented in Fig.~\ref{fig2}f. In contrast to the total phase-shift, $\varphi_{int}$ follows a clear and almost monotonic behavior which shows a hysteresis in the back and forth sweep direction (blue and red curves of Fig.~\ref{fig2}f). $\varphi_{int}$ saturates at $|B_{y}| \gtrsim \SI{15}{\milli\tesla}$ in the two asymptotic limits with total phase drop of $\sim \pi$. Furthermore, during the first magnetization of the SQUID, a clear curve resembling the initial magnetization curve of a ferromagnet has been observed (see Fig.~S2 in the Supplementary Information), confirming the ferromagnetic nature of the impurity ensemble. 

\begin{figure*}[ht]
\includegraphics[center]{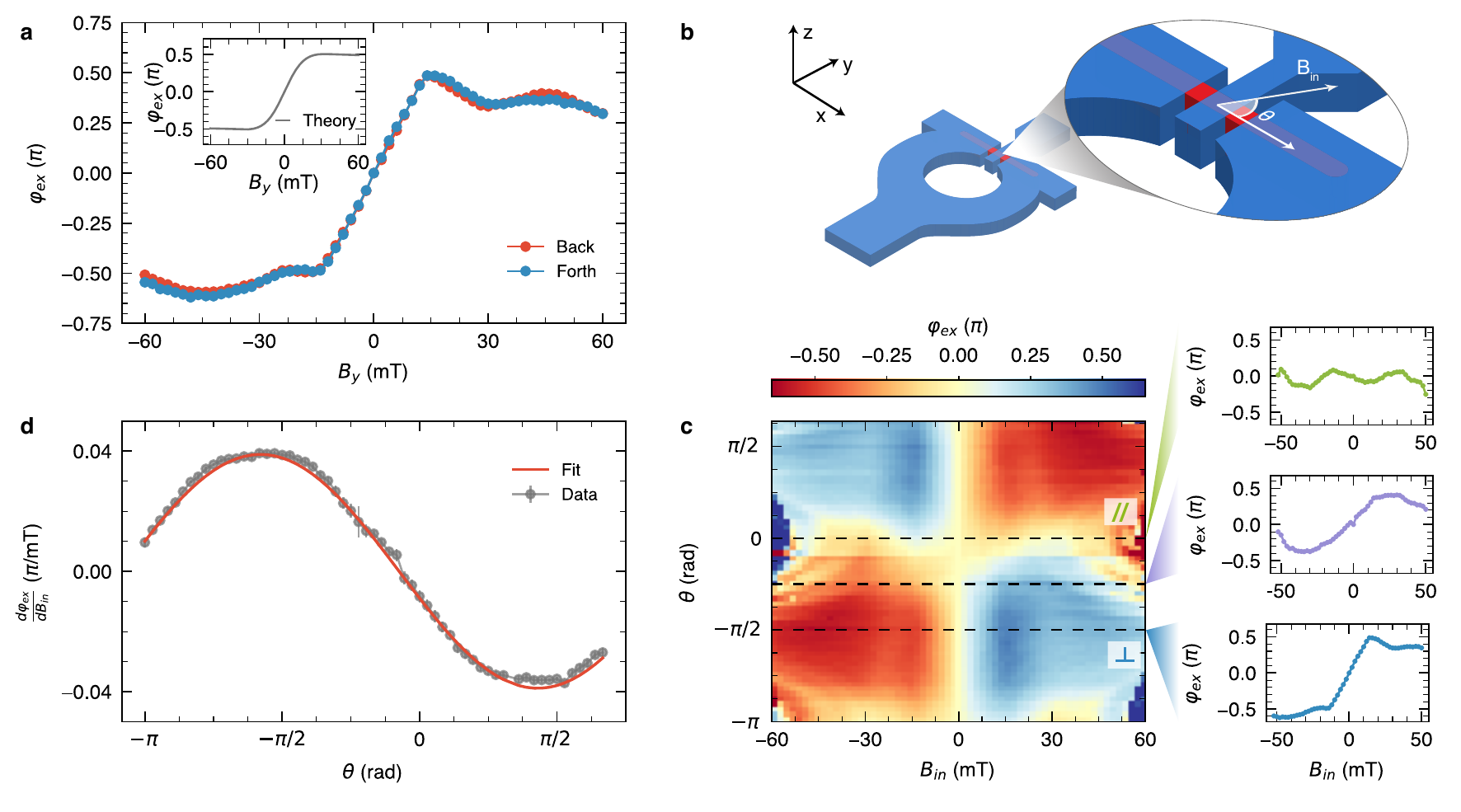}
\caption{\textbf{Vectorial symmetry of the anomalous phase $\varphi_0$.} 
\textbf{a,} Dependence of the extrinsic anomalous phase $\varphi_{ex}$ on $B_y$. It results in an odd symmetry and non-hysteretic back and forth sweeps (blue and red traces). Inset: The $\varphi_{ex}(B_y)$ dependence obtained from the theoretical model (see Section V of the Supplementary Information for details). 
\textbf{b,} Sketch of the interferometer with the reference axes of the in-plane magnetic field ($B_{in}$) and the angle $\theta$ with respect to the nanowire axis. 
\textbf{c,} Dependence of the anomalous phase $\varphi_{ex}$ on $\theta$ and $B_{in}$. The single traces on the right show the behavior of $\varphi_{ex}$ for a longitudinal $\theta = 0$ (green curve), canted $\theta = \pi/4$ (violet curve) and orthogonal field $\theta = \pi /2$ (blue curve).
\textbf{d,} $d \varphi_{ex}/d B_{in}$ versus $\theta$ together with a sinusoidal fit (red curve) from Eq.~\eqref{eq_lowB}. The slope has been evaluated by a linear fit of the data in \textbf{b} for $|B_{in}|< \SI{10}{\milli\tesla}$. The error bar is the $1\sigma$ standard error of the fit. All the data were recorded at $\SI{50}{\milli \kelvin}$ of bath temperature.}
\label{fig3}
\end{figure*}

We now analyze the \emph{extrinsic} contribution, $\varphi_{ex}$, to the phase shift which stems directly from the external $B$-field. Due to the additive nature of the anomalous phase in the exchange field, $h$ in Eq.~\eqref{eq_FL}, it is possible to extract from $\varphi_{tot}$ the extrinsic contribution $\varphi_{ex} = \varphi_{tot} - \varphi_{int}$. This is depicted in Fig.~\ref{fig3}a, where the evolution of $\varphi_{ex}$ in $B_y$ is shown. Here, the agreement between the back (blue) and forth (red) traces in $B_y$ demonstrates implicitly the absence of any hysteresis ensuring the complete extraction of the intrinsic contribution. Notice also that the behavior of $\varphi_{int}$ and $\varphi_{ex}$ in the magnetic field is \emph{opposite} in sign as expected from the competition between the exchange interactions induced by the Kondo antiferromagnetic coupling and a Zeeman field, then further supporting our assumptions. The dependence $\varphi_{ex}(B_y)$ is characterized by a linear increase at low magnetic fields ($|B_{y}|<\SI{15}{\milli\tesla}$) up to a maximum phase-shift of $\pm \pi/2$. Remarkably, our measurement reveals the \emph{odd} parity of the anomalous phase with respect to the magnetic field, one of the main symmetry hallmarks of this effect~\cite{bergeret_theory_2015,buzdin_direct_2008}. This parity is the consequence of the odd parity of the free energy $F_L$ with respect to the exchange field. At higher fields non-linearities appear suggesting a non-trivial evolution of $\varphi_{ex}$ in the magnetic field. In order to understand this behavior we have modelled our $\varphi_0$-junction setup by a lateral junction treated within the quasi-classical approach presented in Ref.~\cite{bergeret_theory_2015} (see Section V in the Supplementary Information). The resulting $\varphi_0$ obtained from the above model, is shown in the inset of Fig.~\ref{fig3}a. It nicely reproduces the main features of $\varphi_{ex}$: the linear dependence at small magnetic fields and the saturation at larger ones. Notice that, within the scale of the magnetic field applied in the experiment, the  field dependence of the anomalous phase looks as it saturates at the value close to $\pi/2$. This value is however non-universal and depends on the characteristics of the nanowire. Moreover, if larger values of $B_{in}$ could be reached, the anomalous phase of each junction would increase up to the universal plateau at $\pi$, as expected also for planar junctions~\cite{bergeret_theory_2015}.  

At small in-plane fields ($B_{in}=\sqrt{B_x^2+B_y^2}$) the model leads to a simple expression for the anomalous phase: 
\begin{equation}
\varphi_{ex} \simeq C_1 \alpha^3 B_{in} \sin{(\theta)} + O(B_{in}^3) ,
\label{eq_lowB}
\end{equation}
where $\theta$ is the angle between the field and the nanowire axis, $C_1$ is a parameter dependent on the temperature and the microscopic details of the JJ and $O$ is the polynomial asymptotic notation. By using typical values of the parameters for the InAs/Al junction we obtained a $C_1 \approx \SI{0.04}{\pi\per\milli\tesla}$, in very good agreement with the experimental data (see Section V in the Supplementary Information).

The odd symmetry of the anomalous phase dictated by the triple product in Eq.~\eqref{eq_FL} can be further investigated by measuring $\varphi_{ex}$ over all the directions of the in-plane magnetic field. Figure~\ref{fig3}c shows the full dependence of $\varphi_{ex}$ on the angle $\theta$ (see sketch in Fig.~\ref{fig3}b). As predicted from Eq.~\eqref{eq_FL}, the phase-shift is very small for fields along the nanowire axis ($\theta = 0$, green trace in Fig.~\ref{fig3}c), showing the maximum slope for the orthogonal magnetic field ($\theta = \pi/2$, blue trace in Fig.~\ref{fig3}c). The odd symmetry manifests clearly as well in the slope $\frac{\partial \varphi_{ex}}{ \partial B_{in}}$ in the low-field limit (Fig.~\ref{fig3}d). The latter is perfectly fitted with a sinusoidal function of $\theta$ in agreement with Eq.~\eqref{eq_lowB} (red trace in Fig.~\ref{fig3}d). 

In summary, our results demonstrate the implementation of a quantum phase battery. This quantum element, providing a controllable and localized phase-bias, can find key applications in different quantum circuits such as energy tuner for superconducting flux~\cite{pita-vidal_gate-tunable_2019} and hybrid~\cite{larsen_semiconductor-nanowire-based_2015} qubits, or persistent multi-valued phase-shifter for superconducting quantum memories~\cite{gingrich_controllable_2016,guarcello_cryogenic_2019} as well as superconducting rectifiers~\cite{reynoso_spin-orbit-induced_2012}. Moreover, the magnetic control over the superconducting phase opens new avenues for advanced schemes of topological superconducting electronics~\cite{virtanen_majorana_2018} based on InAs JJs~\cite{szombati_josephson_2016,tiira_magnetically-driven_2017}. The weak control over the density of unpaired spins makes our proof-of-concept device difficult to reproduce in a massive reliable process. Further technological improvements can be envisioned by a controlled doping of the wires with magnetic impurities~\cite{martelli_manganese-induced_2006} or by the inclusion of a thin epitaxial layer of a ferromagnetic insulator, like EuS~\cite{strambini_revealing_2017}, as recently integrated in similar nanowires~\cite{liu_semiconductor_2019}. 

\section*{Acknowledgement}
The work of E.S. was supported by a Marie Curie Individual Fellowship (MSCA-IFEF-ST No.660532-SuperMag). E.S., N.L and F.G acknowledge partial financial support from the European Union's Seventh Framework Programme (FP7/2007-2013)/ERC Grant No. 615187- COMANCHE. E.S., A.I., O.D., N.L, F.S.B. and F.G were partially supported by EU's Horizon 2020 research and innovation program under Grant Agreement No. 800923 (SUPERTED). L.S and V. Z. acknowledge partial support by the SuperTop QuantERA network and the FET Open And QC. I.V.T, C.S.F., and F.S.B., acknowledge financial support by the Spanish Ministerio de Ciencia, Innovacion y Universidades through the Projects No. FIS2014-55987-P, FIS2016-79464-P and No. FIS2017-82804-P and by the grant “Grupos Consolidados UPV/EHU del Gobierno Vasco” (Grant No. IT1249-19). A.B. acknowledges the CNR-CONICET cooperation program “Energy conversion in quantum nanoscale hybrid devices,” the SNS-WIS joint laboratory QUANTRA, funded by the Italian Ministry of Foreign Affairs and International Cooperation and the Royal Society through the international exchanges between the United Kingdom and Italy (Grant No. IEC R2192166).

\section*{Author contribution}
E.S. A.I. and O.D. performed the experiment and analyzed the data. R.C.,C.S.F.,C.G.,I.V.T., A.B. and F.S.B. provided theoretical support. M.R., N.L. and O.D. fabricated the phase battery on the InAs nanowires grown by V.Z. and L.S.. E.S. conceived the experiment together with F.G. that supervised the project. E.S., A.I., I.V.T.and F.S.B. wrote the manuscript with feedback from all authors.


\bibliographystyle{style}
\bibliography{bibliography.bib}


\clearpage
\onecolumngrid
\setcounter{figure}{0}
\setcounter{equation}{0}
\setcounter{section}{0}
\renewcommand\thefigure{S\arabic{figure}}
\renewcommand\theequation{S\arabic{equation}}
\section*{\large{S\MakeLowercase{upplementary} I\MakeLowercase{nformation}}}

\section{Device fabrication}
Hybrid proximity DC SQUIDs devices were fabricated starting from gold catalyzed $n$-doped InAs nanowires with typical length of $\SI{1.5}{\micro \meter}$ and a diameter of $\sim $ $\SI{85}{\nano \meter}$ grown by chemical beam epitaxy~\citeSI{gomes_controlling_2015}.
The $n$-doping was obtained with Se~\citeSI{wallentin_doping_2011} and the metalorganic precursors for the nanowire growth was trimethylindium (TMIn), tertiarybutylarsine (TBAs) and ditertiarybutylselenide (DTSe), with line pressures of 0.6, 1.5 and 0.3 Torr respectively.
Nanowires were drop-casted onto a substrate consisting of $\SI{300}{\nano \meter}$ thick SiO$_2$ on $p$-doped Si. Afterwards, a $\SI{280}{\nano \meter}$-thick layer of positive-tone Poly(methyl methacrylate) (PMMA) electron beam resist was spun onto the substrate. The devices were then manually aligned to the randomly distributed InAs nanowires and patterned by means of standard electron beam lithography (EBL) followed by electron beam evaporation (EBE) of superconducting Ti/Al ($\SI{5/100}{\nano \meter}$) electrodes. Low-resistance ohmic contacts between the superconducting leads and the InAs nanowires were promoted by exposing the InAs nanowire contact areas to a highly diluted ammonium polysolfide (NH$_4$)S$_x$ solution, which selectively removes the InAs native oxide and passivates the surface, prior to EBE. The fabrication process was finalized by dissolving the PMMA layer in acetone.

From transport characterization on similar wires and normal metal electrodes~\citeSI{iorio_vectorial_2019}, we estimate a typical electron concentration $n \simeq \SI{2e18}{\cm^{-3}}$ and mobility $\mu \simeq \SI{1200}{\cm^2/\volt\second}$. The corresponding Fermi velocity $v_F$, mean free path $l_e$ and diffusion coefficient $D = v_F l_e/3$, are evaluated to be $v_F \simeq \SI{2e6}{\m/\s}$, $l_e \simeq \SI{30}{\nm}$ and $D \simeq \SI{200}{\cm^2/s}$.

\section{Kondo resistance $R(T)$}
\begin{figure}[h]
\centering
\includegraphics{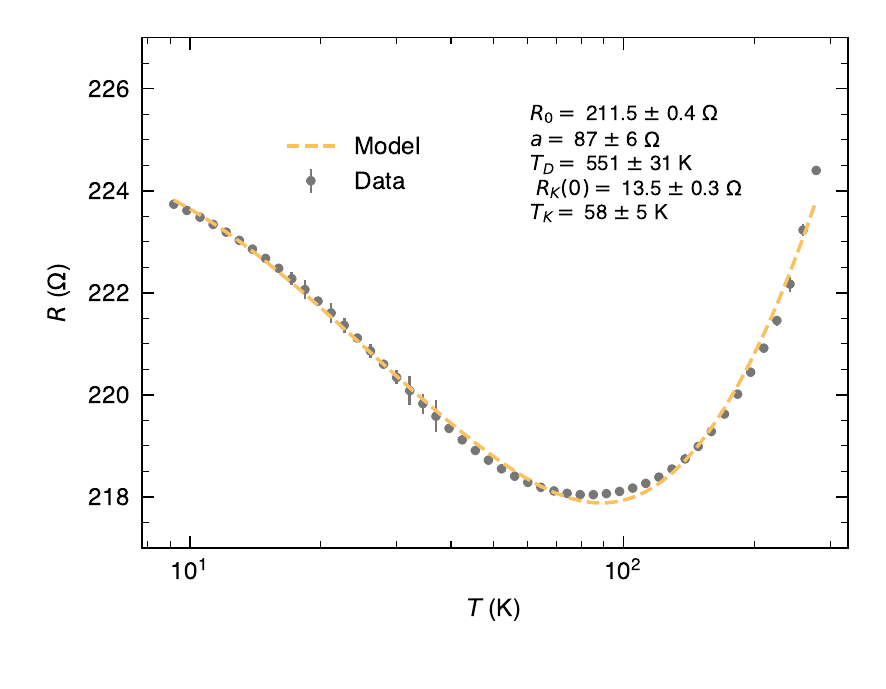}
\caption{\textbf{Kondo upturn of the InAs nanowire.} Resistance versus temperature $R(T)$ of one of the devices measured during the cooldown of the refrigerator showing a clear increase of resistance for temperature below 80~K. This behaviour is consistent with the Kondo scattering model as demonstrated by the good fit of the $R-T$ data with the model in Eq.~\ref{eq:kondo}.}
\label{fig_kondo}
\end{figure}

To quantify the amount of unpaired spins in our system the temperature dependence of the normal-state resistance $R(T)$ has been studied in the range \SIrange{10}{300}{\kelvin}. The data exhibits an upturn at $T\sim\SI{80}{\kelvin}$ (see Fig.~\ref{fig_kondo}) suggesting a Kondo scattering mechanisms between the free electrons and the unpaired spins in the weak-links. Since the InAs nanowires were synthesized without incorporating any magnetic impurity (to the best of our knowledge, Se doping cannot provide by itself any magnetism), we conjecture that unpaired spins are originated from oxides states at the nanowire surface, in analogy with what is observed in metallic nanowires~\citeSI{rogachev_magnetic-field_2006,sapkota_observations_2018}.
Indeed, the $R(T)$ of a diluted magnetic alloy follows the universal non-monotonic relation~\citeSI{kondo_resistance_1964} 
\begin{equation}
R(T) = R_0 + R_{el-ph}(T) + R_K(T),
\end{equation}
where $R_0$ is the residual resistance while $R_{el-ph}(T)$ and $R_K(T)$ are the contribution given respectively by the electron-phonon and the Kondo scattering. The temperature dependence of the former can be expressed according to the Bloch-Gruneisen model as~\citeSI{PhysRevB.74.035426}
\begin{equation}
R_{el-ph}(T) = a\left(\frac{T}{\theta_D}\right)^5 \int_0^{\theta_D/T} \frac{x^5}{(e^{x}-1)(1-e^{-x})}\text{d}x.
\end{equation}
For the Kondo contribution many analytical approximations are available according to the range of temperature investigated. In the full range of temperature the exact solution exist from the numerical renormalization group theory (NRG). In the following we use an empirical fitting function derived as an analytical approximation of the NRG given by~\citeSI{goldhaber-gordon_kondo_1998,Parks1370, mallet_scaling_2006,costi_kondo_2009}
\begin{equation}
R_K^{NRG} = R_K(0) \left ( \frac{T_K'^2}{T^2+T_K'^2} \right )^s,
\label{eq:kondo}
\end{equation}
with $T_K'$ related to the actual Kondo temperature $T_K$ by $T_K'=T_K/(2^{1/s}-1)^{1/2}$. Note that Eq.~\ref{eq:kondo} is defined such that $R_K(T_K)=R_K(0)/2$ and the parameter $s$ is fixed to $s=0.22$ as expected for a spin $1/2$ impurity. In Fig.~\ref{fig_kondo} we show the fit of the experimental data with Eq.~\ref{eq:kondo}, from which we extract a Kondo temperature $T_K = 58 \pm 5$ K, a residual magnetic impurity resistance $R_K(0)=13.5 \pm 0.3~\Omega$, a coefficient $a=87 \pm 6~\Omega$ and a Debye temperature $\theta_D = 551 \pm 31$ K. From $R_K(0)$ is possible to estimate the density of unpaired spin, that form the Hamann expression of the residual Kondo resistance in the unitary limit is given by~\citeSI{hamann_new_1967}
\begin{equation}
R_K(0) = \frac{L}{A} \frac{4\pi c \hbar}{n k_F e^2},
\end{equation}
with $L\sim \SI{80}{\nm}$ junction lenghts, $A \sim \pi r^2$ with $r \sim \SI{45}{\nm}$ nanowire cross-sectional area, $k_F$ Fermi wavevector and $n$ electron carrier density, from which we estimate the density of magnetic impurities $c \simeq \SI{1.36e17}{\cm^{-3} }$. This corresponds to a concentration of $\sim$ 4 ppm ($= c /n_{InAs}$, with the InAs atomic density $n_{InAs} = \SI{3.59e22}{\cm^{-3}}$) of unpaired spins in the InAs nanowire.

\section{First ``magnetization'' curve}
\begin{figure*}[t]
\centering
\includegraphics[center]{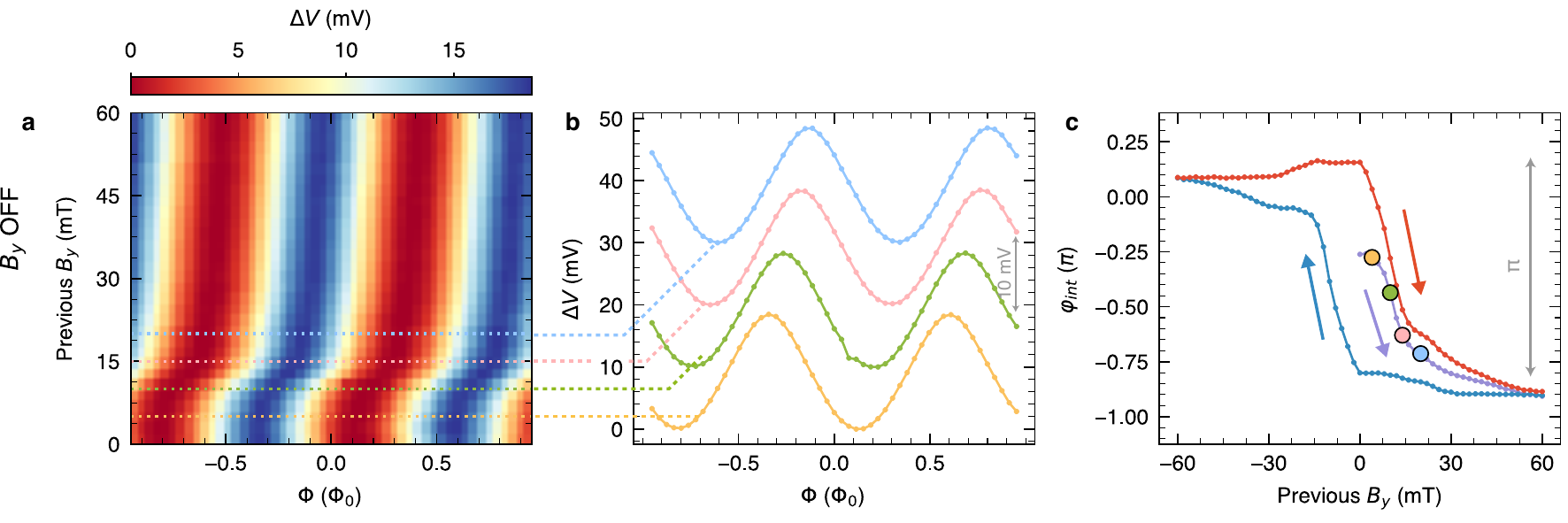}
\caption{\textbf{First magnetization curve of $\varphi_{int}$.} \textbf{a}, Color plot of the voltage drop $\Delta V (\Phi)$ measured at $B_y = 0$ after the magnetic field was swept to the values shown on the $y$-axis straight after the thermal cycle. \textbf{b}, Selected traces $\Delta V (\Phi)$ corresponding to the cuts in \textbf{a}. \textbf{c}, Intrinsic phase shift $\varphi_{int}$ extracted from \textbf{b}, showing the first polarization of the unpaired spins (violet curve) and the hysteresis loop followed in the subsequent back and forth sweeps of $B_y$, blue and red curves, respectively.
}
\label{fig:SI_first_magnetization}
\end{figure*}

The persistent hysteretic loops of the $\varphi_0$-shift, shown in Fig.~1d-f of the main text, are consistent with the presence of a ferromagnetic background of unpaired spin. To support this hypothesis, we show in Fig.~\ref{fig:SI_first_magnetization}a and b the first magnetization curve of this spin ensemble measured in the same device.
Initially, the magnetization of the sample is lifted by warming the system above 3 K. Then, the SQUID voltage drop $\Delta V (\Phi)$ is measured in the absence of the in-plane magnetic field $B_y$ which is gradually turned on thus polarizing the unpaired spins. The resulting $\Delta V (\Phi)$ shows no shifts at low $B_y$ while, only above $\SI{5}{\milli \tesla}$ a clear shift is generated. The resulting $\varphi_{int}$ extracted by fitting $\Delta V (\Phi)$ is shown in the violet curve of Fig.~\ref{fig:SI_first_magnetization}c. By reversing $B_y$ the $\varphi_{int}$ then evolves with the typical hysteretic curve of a ferromagnetic system (blue and red curves in Fig.~\ref{fig:SI_first_magnetization}c).

\section{SQUID with anomalous Josephson junctions}
The critical current of a SQUID interferometer can be evaluated from the CPR of the two JJs forming the interferometer. 
Using a sinusoidal CPR, the currents through the two junctions can be written as
\begin{equation}
i_1 = I_{c} \sin\left[(\varphi_C-\varphi_L)+\varphi_0^{(1)}\right]  \qquad i_2 = I_{c} \sin\left[(\varphi_C-\varphi_R)+\varphi_0^{(2)}\right],
\label{eq:1-2}
\end{equation}
where $\varphi_L,\varphi_C,\varphi_R$ are the left,central and right superconducting phases and $I_{c}$ is the critical current of each JJ.

The supercurrent of the SQUID is the sum of the two contributions ($I_s = i_1 + i_2,$) and, with the constraint on the superconducting phases of the flux quantization 
\begin{align}
(\varphi_L -\varphi_C)  + (\varphi_C - \varphi_R) +2\pi\frac{\Phi}{\Phi_0} 
&= 2\pi~\text{(mod n)},
\label{eq:flux_quantization}
\end{align}
it has the form
\begin{align}
I_s &= 2I_C \sin(\delta_0) \cos\left[\frac{1}{2}\left(2\pi\frac{\Phi}{\Phi_0} +\varphi_{tot}\right)\right],
\end{align}
\begin{figure}
\centering
\includegraphics{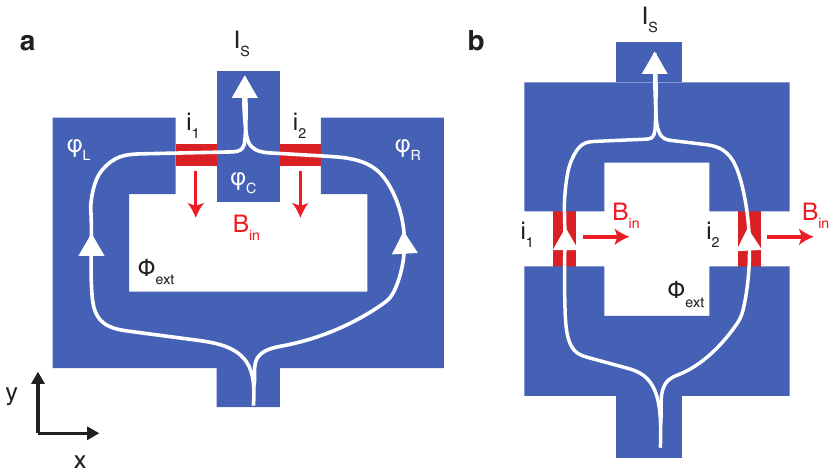}
\caption{\textbf{Comparison between SQUID geometries.} The SQUID geometry employed in this work \textbf{a} allows a simple readout of the anomalous phase $\varphi_{tot}$ generated by a uniform magnetic field $B_{in}$ since the opposite direction of the supercurrents in the two JJs. This would not be possible in a conventional geometry as the one shown in \textbf{b}.}
\label{fig:squid_geometry}
\end{figure}
where $\delta_0 =\varphi_C- \tfrac{\varphi_L+\varphi_R}{2} +\tfrac{(\varphi_0^{(2)}+\varphi_0^{(1)})}{2}$ and $\varphi_{tot}=\varphi_0^{(2)}-\varphi_0^{(1)}$ is the total anomalous phase built in the interferometer. With the geometry depicted in Fig.~\ref{fig:squid_geometry}a, the two junctions experience the same in-plane magnetic field orientation but the supercurrents flow in opposite directions resulting in $\varphi_0^{(1)}=-\varphi_0^{(2)}=\varphi_0$ and $\varphi_{tot}=2\varphi_0$. The stable state configuration of the SQUID is achieved by minimizing the total Josephson free energy obtained at $\delta_0=\pi/2$, and then the maximum sustainable supercurrent results to be
\begin{equation}
I_{S}(\Phi) = 2I_C\left|\cos\left(\pi\frac{\Phi}{\Phi_0} +\frac{\varphi_{tot}}{2}\right)\right|.
\end{equation}
It follows that in absence of magnetic flux, the maximum supercurrent is reduced by a factor $\sim |\cos(\varphi_{tot}/2)|$ compared to the non-anomalous case as a consequence of the anomalous supercurrent already present in the interferometer. In a more conventional geometry as the one showed in Fig.~\ref{fig:squid_geometry}b, the anomalous phases acquired by the two junctions would be the same $\varphi_0^{(1)}=\varphi_0^{(2)}=\varphi_0$, making impossible its detection in the phase-to-current readout employed in the present work.

\section{Lateral $\varphi_0$-junction}
\label{sec:theory}
The origin of the anomalous phase $\varphi_0$ is the singlet-triplet conversion mediated by the spin-orbit coupling (SOC), which in the normal state corresponds to the charge-spin conversion~\citeSI{bergeret_theory_2015}. 
The calculations of the anomalous Josephson current in $\varphi_0$-junctions have been done for ideal planar S-N-S junctions, in which the superconducting electrodes and the normal region with SOC are separated by sharp boundaries~\citeSI{bergeret_theory_2015,konschelle_theory_2015,buzdin_direct_2008}, where the singlet-triplet coupling takes place only at N region. As shown in Ref.~\citeSI{bergeret_theory_2015}, this assumption leads to a monotonically increase of the anomalous phase $\varphi_0$ as a function of the applied magnetic field, which contrasts with curves extracted from our experiment (see Fig.~3a in the main text). 
It is however clear that our experimental setup (Fig.~1 in the main text) differs from an ideal S-N-S junction. Indeed, in each junction, the superconducting leads are covering part of the wires over distances larger than the coherence length. This means that the SOC, and hence the spin-charge conversion, is also finite in the portion of the wire covered by the superconductor. As we show in this section, this feature is essential to understand the experimental findings; in particular, the dependence to $\varphi_0$ from the external magnetic field. In this calculation, we focus on the dependence of $\varphi_0$ on the $y$ direction of the field, i.e., $B_{in}$ at $\theta = \pi/2$ (see Figs.~3a and b in the main text). 

\begin{figure*}
 \centering
 \includegraphics[center]{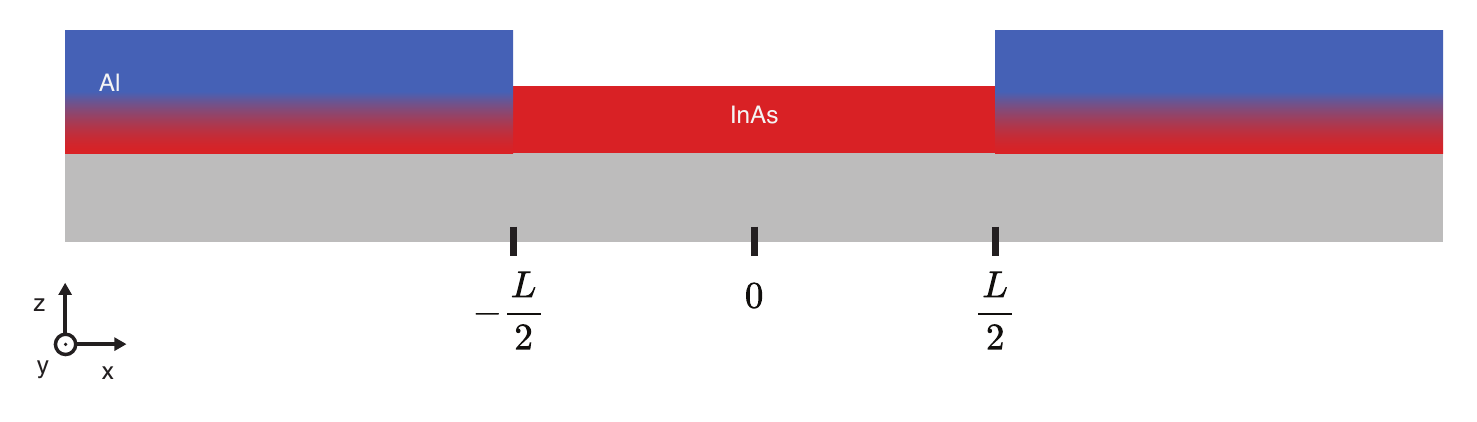}
 \caption{\textbf{Schematic view of the S-N-S junctions.} At $|x|>L/2$, the InAs nanowire (red region) is partially covered by the Al superconducting leads (dark blue regions). The gray region corresponds to the substrate. This schematic view corresponds to the $z$-$x$ plane of Fig.~3b (top panel) in the main text.}
 \label{fig_sketch}
\end{figure*}

To be specific, we consider the junction sketched in Fig.~\ref{fig_sketch}. We assume an infinite diffusive quasi-one dimensional nanowire along the $x$-axis, which is partially covered by two semi-infinite Al superconducting leads at $x<L/2$ and $x>L/2$. We assume, for simplicity, that the proximity effect is weak and that the wire is diffusive. In such a case, the condensate function, which determines the Josephson current, obeys the linearized Usadel equation, which results in two coupled differential equations for the singlet and triplet components as shown in Ref.~\citeSI{bergeret_theory_2015}. 
Because the wire lies on a substrate plane, the system has an uniaxial asymmetry in the $z$ direction perpendicular to the substrate (see Fig.~\ref{fig_sketch}). In the presence of SOC, this allows for a gradient singlet-triplet coupling generated by a differential operator of the form $C^a_k\partial_k\sim(\hat z\times\nabla)^a$, which converts a scalar (the singlet) into a pseudovector (the triplet) and vice-versa~\citeSI{konschelle_theory_2015,bergeret_theory_2015}. We consider the case when the external field is applied in the $y$ direction, and hence, the superconducting condensate function has the form $f=f_s+if_t{\rm sgn}\omega\sigma^y$, where $f_{s,t}$ are the singlet and triplet components and $\omega$ the Matsubara frequency. The linearized Usadel equation reads: 
\begin{equation}
\label{eq:3D_usadel}
\begin{split}
&\frac{D}{2}\nabla^2 f_s - |{\omega}| f_s + \left( h - iD \kappa_{\rm{sc}} \partial_x \right) f_t = 0 \; , \\
&\frac{D}{2}\nabla^2 f_t - |{\omega}| f_t - \left( h - 2i D\kappa_{\rm{sc}} \partial_x \right) f_s = 0 \; . \\
\end{split}
\end{equation}
Here $D$ is the diffusion coefficient and $h={\mu_{\rm{B}} g_{\rm{s}}} B_{\rm{in}} / 2$ is the Zeeman field. The last term in both equations describes the spin-charge conversion due to the SOC. It is proportional to 
the effective inverse length $\kappa_{\rm{sc}}$ and the spatial variation of the condensate in the direction of the wire axis. The form of this term is determined by the uniaxial anisotropy of the setup in combination with the fact that we assume that the field is applied only in $y$ direction. 

Equation~\eqref{eq:3D_usadel} is written for the full 3D geometry. To obtain an effective 1D Usadel equation, we integrate Eq.~\eqref{eq:3D_usadel} over the wire cross-section and use boundary conditions imposed on the condensate function at the surface of the wire. In the part of the wire which is covered by the superconductor, the interface between the wire and the superconductor is described by the linearized Kupryianov-Lukichev boundary condition:
\begin{equation}
\left.\partial_x f_s \right|_{\text{InAs/Al}}=\gamma f_{\rm{BCS}}e^{i\phi}, 
\end{equation}
where $\gamma$ is a parameter describing the InAs/Al interface, $f_{\rm{BCS}} = \Delta / \sqrt{\omega^2 + \Delta^2}$ is the BCS bulk anomalous Green's function in the superconducting leads, and $\phi$ is the phase of the corresponding lead. In the uncovered parts of the wire, we impose a zero current flow which corresponds to 
$\left.\partial_x f_{s/t}\right|_{\text{InAs/vac.}}=0$. The integration of Eq.~\eqref{eq:3D_usadel} over the cross-section of the wire results in two coupled equations for the singlet and triplet components:
\begin{equation}
\label{eq:1D_usadel}
\begin{split}
&\partial_x^2 f_s - \kappa_{\omega}^2 f_s + \left( \kappa_{h}^2 - 2i \kappa_{\rm{sc}} \partial_x \right) f_t = S(x) \; , \\
&\partial_x^2 f_t - \kappa_{\omega}^2 f_t - \left( \kappa_{h}^2 - 2i \kappa_{\rm{sc}} \partial_x \right) f_s = 0 \; , \\
\end{split}
\end{equation} 
with 
\begin{equation}
\begin{split}
S(x) = \gamma f_{\rm{BCS}} \left[ \Theta \left( x - \frac{L}{2} \right) e^{i\frac{\varphi}{2}} + \Theta \left( - x - \frac{L}{2} \right) e^{-i\frac{\varphi}{2}} \right] \; ,
\end{split}
\end{equation}
 $\kappa_{\omega}^2 = \frac{2|\omega|}{D}$, $\kappa_h^2 = \frac{2h}{\hslash D}$, and $\varphi$ is the phase difference between these two Al leads.
After a cumbersome but straightforward procedure, we solve Eq.~\eqref{eq:1D_usadel} for continue and finite $f_{s,t}$. 
From the knowledge of the singlet and triplet components one determines the Josephson current as follows~\citeSI{bergeret_theory_2015}:
\begin{equation}
\begin{split}
&j(x) = \frac{\pi \sigma_{\rm{D}} T}{e} \sum_{\omega} \operatorname{\mathbb{I}m} \lbrace f_s^* \partial_x f_s - f_t^* \partial_x f_t - i \kappa_{\rm{sc}} \left( f_s^* f_t + f_s f_t^* \right) \rbrace\; .
\end{split}
\end{equation}
The resulting current can be written as $j = I_c \sin(\varphi + \varphi_0)$, with the anomalous-phase given by:
\begin{equation}
\label{eq:final}
\begin{split}
\varphi_0 &= \arctan\left\{ \frac{ \sum_{\omega}\operatorname{\mathbb{I}m}\left\{ f_{\rm{BCS}}^2 e^{-q L} \frac{\sinh (\kappa_{\rm{sc}} L) (q^{2} + \kappa_{\rm{sc}}^2) + 2 q\kappa_{\rm{sc}}\cosh (\kappa_{\rm{sc}} L)}{q (q^{2} - \kappa_{\rm{sc}}^2)^2} \right\}}{ \sum_{\omega}\operatorname{\mathbb{R}e}\left\{ f_{\rm{BCS}}^2 e^{-q^* L} \frac{\cosh (\kappa_{\rm{sc}} L) (q^{*2} + \kappa_{\rm{sc}}^2) + 2 q^*\kappa_{\rm{sc}}\sinh (\kappa_{\rm{sc}} L)}{q^* (q^{*2} - \kappa_{\rm{sc}}^2)^2} \right\}} \right\} \; , \\
\end{split}
\end{equation}
where $q^2 = \kappa_{\omega}^2 + \kappa_{\rm{sc}}^2 - i \kappa_{h}^2$. In order to compare with the experimental data, we assume a Rashba-like SOC and use the expression derived in Ref.~\citeSI{bergeret_theory_2015} for the spin-charge coupling parameter, namely $\kappa_{\rm{sc}}=2\tau\alpha^3m^{*2}/\hbar^5$. By using typical values for the parameters of a InAs/Al system: $\xi_0 \simeq \SI{100}{\nano\meter}$, $\Delta \simeq \SI{150}{\micro\eV}$, $m^* = 0.023~m_e$, $T\simeq \SI{25}{\milli\kelvin}$, $g_s \simeq 12$, and $\alpha \simeq \SI{0.24}{\eV\angstrom}$, we find the $\varphi_0(B_{in})$ dependence corresponding to the one shown in Fig.~3a of the main text. We see that our model provides a good qualitative explanation of the two main observed features. Namely, the linear increase of $\varphi_0$ for small fields and a kind of saturation at $\varphi_0 \approx \pm 0.5 \pi$. 

\begin{figure*}[t]
 \centering
 \includegraphics[center]{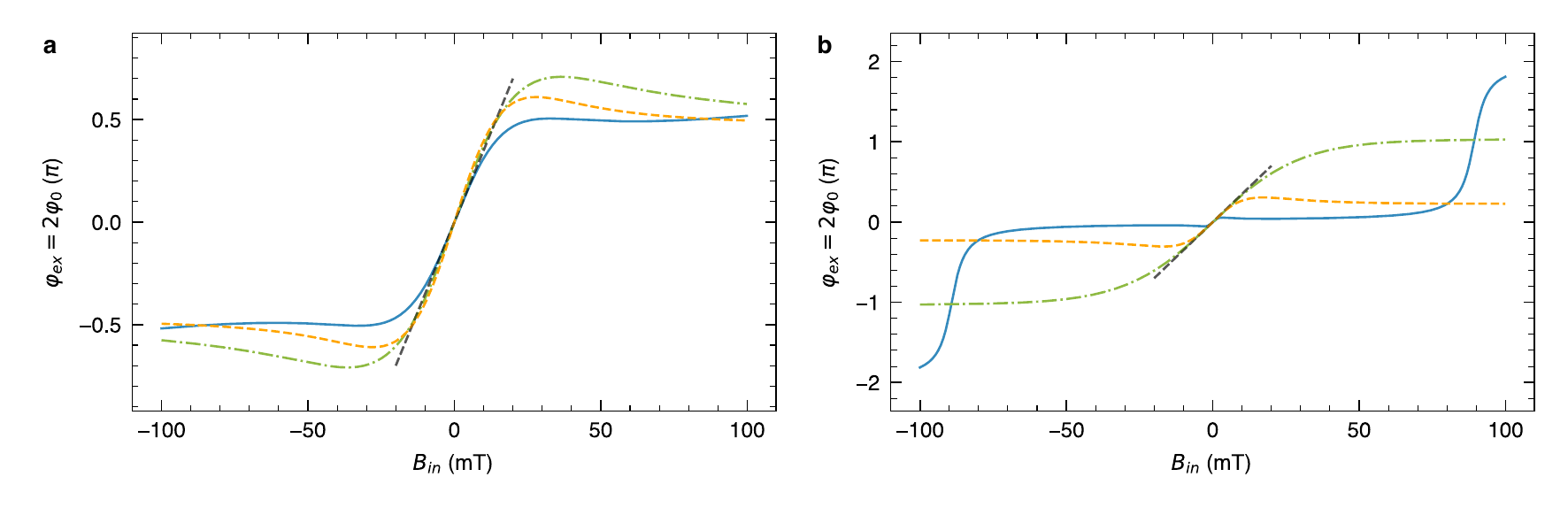}
 \caption{\textbf{Theoretical model of the extrinsic anomalous phase.} The dependence of the extrinsic anomalous-phase on the magnetic field applied in $y$ direction ($B_{in}$ at $\theta = \pi/2$) for a) different temperatures, with $\alpha \simeq  \SI{0.24}{\eV\angstrom}$ and for b) different values of $\alpha$, with $T\simeq \SI{10}{\milli\kelvin}$. In a), the solid blue line coincides with the one shown in Fig.~3a of the main text, with $T\simeq \SI{25}{\milli\kelvin}$ and $g_s \simeq 12$. For the dashed orange line, we choose $T\simeq \SI{10}{\milli\kelvin}$ and $g_s \simeq 5$, and for the dashed-dotted line, $T\simeq \SI{5}{\milli\kelvin}$ and $g_s \simeq 2$. In b), the solid blue line corresponds to $\alpha \simeq \SI{0.1}{\eV\angstrom}$ and $g_s \simeq 37$, the dashed orange line to $\alpha \simeq \SI{0.18}{\eV\angstrom}$ and $g_s \simeq 7$, and the dashed-dotted green line to $\alpha \simeq \SI{0.4}{\eV\angstrom}$ and $g_s \simeq 3$. In both a) and b), the dashed grey line corresponds to the measured slope at low fields.}
 \label{fig:diff_T}
\end{figure*}

In Fig.~\ref{fig:diff_T}, we show different $\varphi_0 (B_{in})$ curves obtained from our general expression~\eqref{eq:final}. Whereas for small fields the experimental slope (dashed grey line) can be obtained from different values of the parameters, the behaviour of $\varphi_0$ at larger fields depends strongly on these parameters.

Indeed, it is important to emphasize that the saturation value at $\varphi_0\approx \pm 0.5 \pi$ is not an universal property of the phase-battery. This value depends on the intrinsic properties of the system. In particular, larger values of the SOI $\alpha$ leads to larger values of $\varphi_{\rm{ex}}$ at values of the field larger than those accessed in the experiment. This is shown in Fig.~\eqref{fig:diff_T}b, where we plot the $\varphi_0 (B_{in})$ dependence for different values of $\alpha$, with $T \simeq \SI{10}{\milli\kelvin}$. As in Fig.~\eqref{fig:diff_T}a, we change the $g_s$ value to maintain the experimental slope in the low-field region.
The linear behavior for the low-field region is shared by all the $\varphi_0 (B_{in})$ curves, as shown in Fig.~\eqref{fig:diff_T}. In this regime, we can thus find the slope value by linearizing Eq.~\eqref{eq:final}:
\begin{equation}
\varphi_{0} \simeq C_1 B_{in} + O(B_{in}^3) ,
\label{eq_lowB_SI}
\end{equation}
with $C_1 \simeq \SI{0.035}{\pi\per\milli\tesla}$, which is in agreement with the value extracted from the experiment.

\section{Trivial mechanisms to induce phase shifts}
Trivial hypotheses, alternative to the anomalous $\varphi_0$ effect, have been also considered for the for the generation of a hysteretic phase shift: trapped magnetic fluxes and Abrikosov vortices.
 
\begin{itemize}
 \item Trapped magnetic fluxes can be observed in superconducting loops with a non negligible ring inductance $L$ and, more precisely, for a screening parameter $\beta_L =\frac{ 2\pi LI_c}{\Phi_0} \gtrsim 1$, with $I_C$ being the critical current of a single junction~\citeSI{clarke_squid_2004}. This indeed can lead to a hysteretic behavior due to the presence of a circulating current in the ring. For our interferometer we estimated $\beta_L \lesssim 10^{-2}$ ($I_C \sim \SI{300}{\nA}$ and $L \sim \SI{10}{\pico\henry}$~\citeSI{dambrosio_normal_2015}) that is very unlikely to induce any magnetic hysteresis. Still, if circulating currents are present, hysteretic jumps should be sharp, periodic and visible even at low $B_z$. The absence of any hysteretic behavior at low magnetic field is further confirmed by the continuous interference patterns shown in Figs.~1e and 1f.
 
\item Abrikosov vortices, also known as fluxons, can be often induced in type-II superconductors – like the thin Al film used in our SQUID devices – when an out-of-plane magnetic field is applied. To avoid vortex intrusion into the ring surface, which might induce a parasitic phase shift, we limit our out-of-plane component to $|B_z| < \SI{0.8}{\milli\tesla}$, which guarantees the absence of any fluxon. Indeed, upon the application of a larger field $B_z \gtrsim (3-4)~\text{mT}$, also in our case abrupt phase shifts appear with a density that increases by increasing $B_z$, as shown in Fig.~\ref{fig:SI_fluxons_Bz}. 
\begin{figure*}[t]
 \centering
 \includegraphics{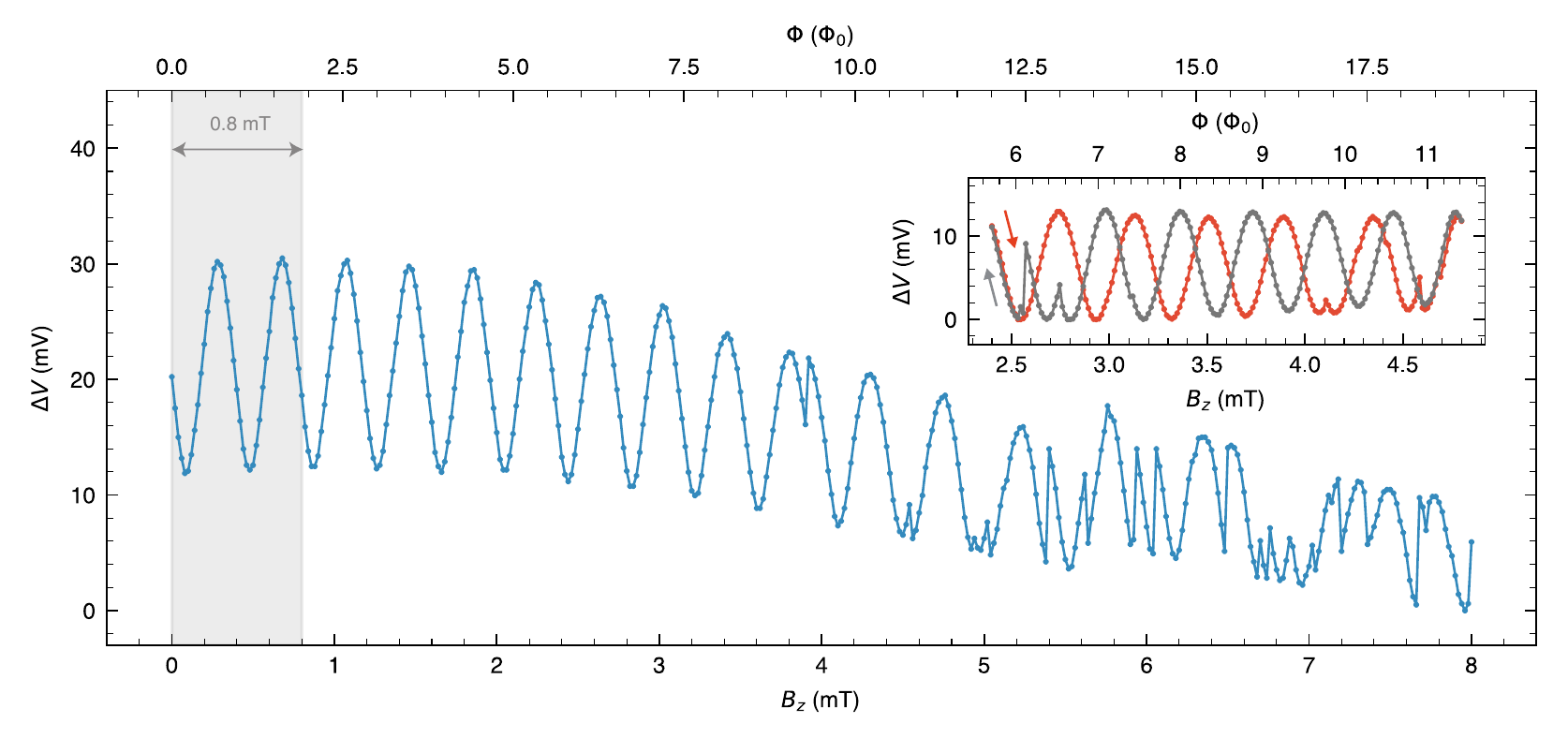}
 \caption{\textbf{Fluxons induced phase shifts at high $B_z$.} Voltage drop $\Delta V(B_z)$ across the SQUID for $I_{sd} = \SI{1}{\micro\ampere}$ versus applied magnetic field $B_z$ up to \SI{8}{\milli\tesla} recorded at $T=\SI{100}{\milli\kelvin}$. The grey area indicated in the plot (corresponding to $0\leq B_z\leq \SI{0.8}{\milli\tesla}$) is the one used to track and evaluate the induced phase shift in our interferometer. For $B_z \gtrsim \SI{3}{\milli\tesla}$ abrupt jumps in the phase start to appear due to trapped fluxons piercing the SQUID area. Inset: back (gray) and forth (red) traces at high $B_z$ in the same conditions as before show an hysteretic behavior which is expected for fluxons pick-up.}
 \label{fig:SI_fluxons_Bz}
\end{figure*}
This is what is expected for fluxons pinning in the Al, i.e., stochastic and abrupt events providing a discrete jump of the phase~\citeSI{PhysRevLett.104.227003}. Notice also the hysteretic behavior expected for fluxon inclusion, which is underlined in the inset of the figure showing a local back and forth measurement. 

With respect to the in-plane magnetic fields, the thickness of the Al film (thinner than the superconducting coherence length) ensures the complete penetration of the magnetic field, and thereby the absence of generated fluxons. This is consistent with the lack of any stochastic shift upon the application of $B_{in}$.
\end{itemize}

\section{Supplementary device measured}
\begin{figure*}[h]
\centering
\includegraphics[center]{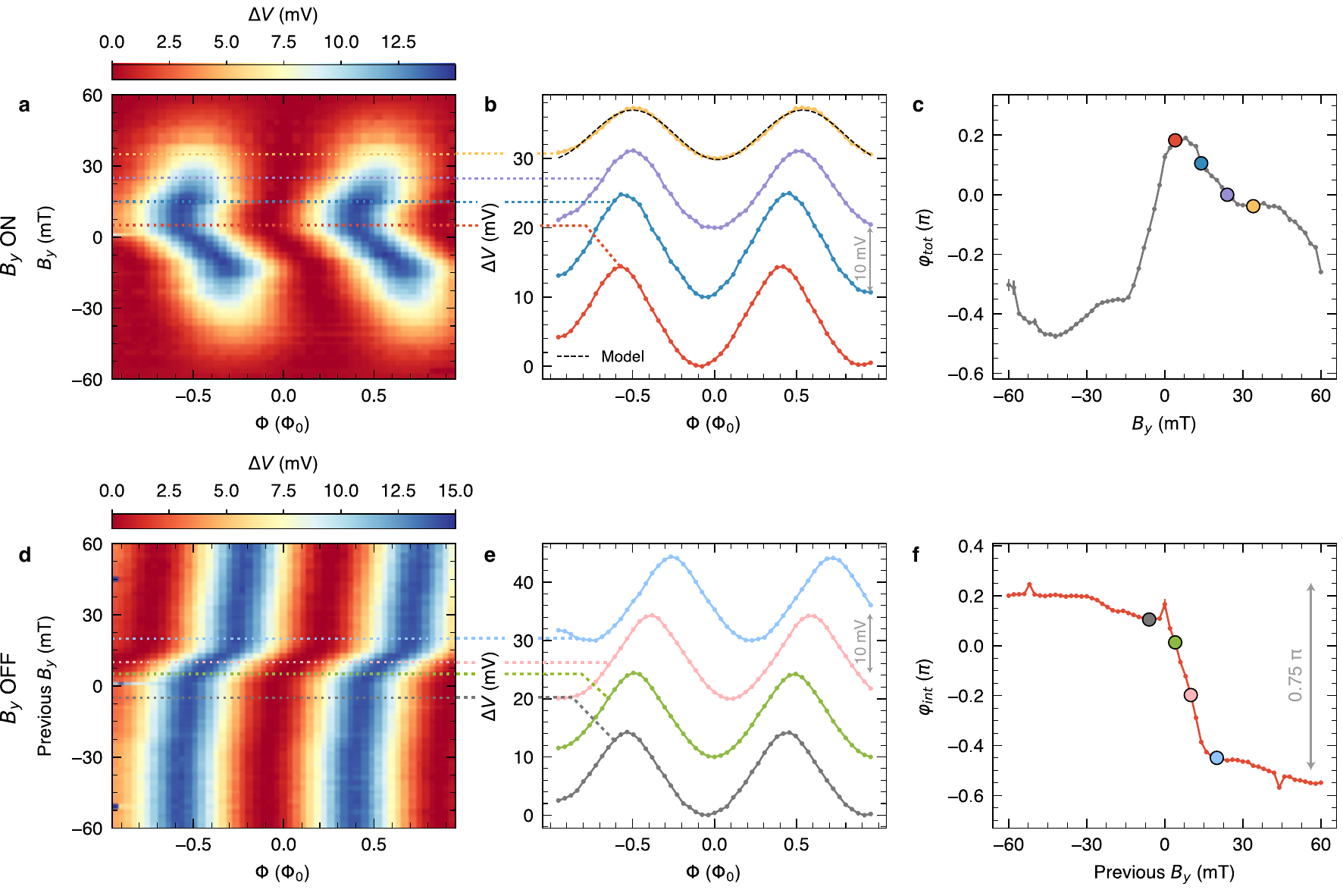}
\caption{\textbf{Charging loops of the Josephson phase battery (second device).} 
\textbf{a,} Voltage drop $\Delta V(\Phi)$ at constant current bias $I=\SI{1}{\micro\ampere}$ versus in-plane magnetic field $B_{y}$ applied orthogonal to the nanowire axis . At large $|B_{y}|$, the amplitude of $\Delta V(\Phi)$ is lowered due to the suppression of superconductivity inside the wire. Each trace is vertically offset for clarity.
\textbf{b,} Selected traces $\Delta V(\Phi)$ extracted from \textbf{a} for different $B_{y}$. Data are vertically offset for clarity. 
\textbf{c,} Extracted phase shift $\varphi_{tot}$ from the curves in \textbf{a}. 
\textbf{d,} Color plot of the persistent voltage drop $\Delta V(\Phi)$ measured at $B_{y}=0$ after the magnetic field was swept to the values shown on the $y$-axis. 
\textbf{e,} Selected traces $\Delta V(\Phi)$ corresponding to the cuts in \textbf{d}. 
\textbf{f,} Intrinsic phase shift $\varphi_{int}$ extracted from \textbf{d}. $\varphi_{int}$ stems from the ferromagnetic polarization magnetic impurities. All data were recorded at $\SI{30}{\milli \kelvin}$ of bath temperature.}
\label{fig:SI_device_extra1}
\end{figure*}

\begin{figure}[H]
\centering
\includegraphics[center]{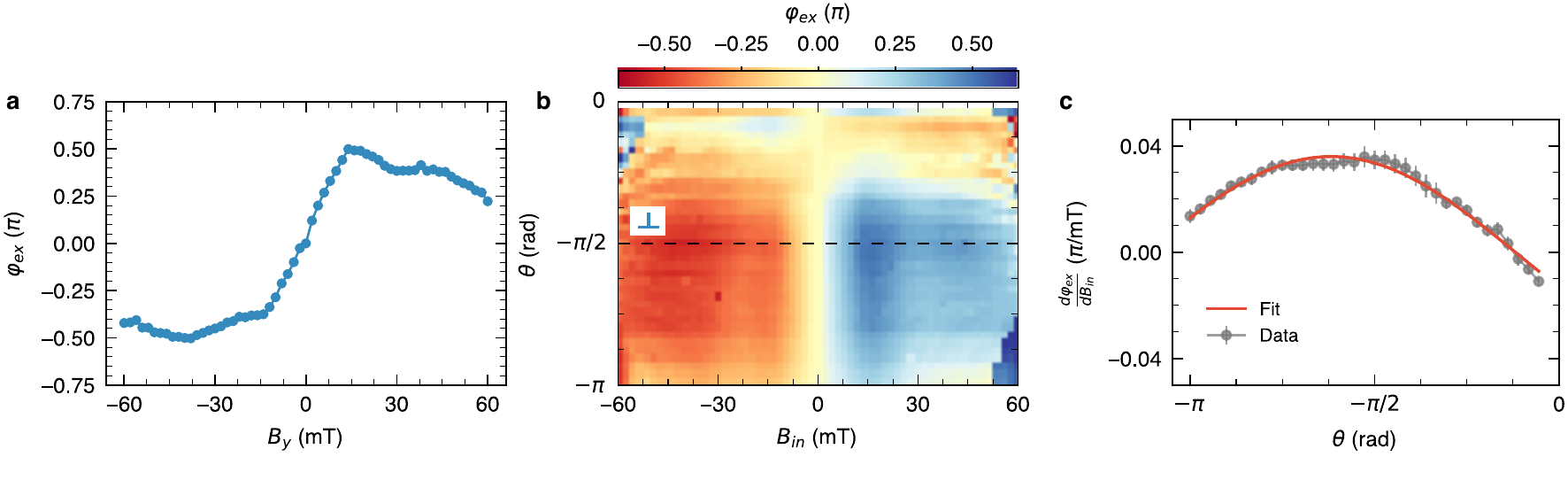}
\caption{\textbf{Vectorial symmetry of the anomalous phase $\varphi_0$ (second device).} 
\textbf{a,} Dependence of the extrinsic anomalous phase $\varphi_{ex}$ on $B_y$. 
\textbf{b,} Evolution of the anomalous phase $\varphi_{ex}$ on $\theta$ and $B_{in}$. 
\textbf{c,} $d \varphi_{ex}/d B_{in}$ versus $\theta$ together with a sinusoidal fit (red curve) from Eq.~\eqref{eq_lowB_SI}. The slope has been evaluated by a linear fit of the data in \textbf{b} for $|B_{in}|< \SI{10}{\milli\tesla}$. The error bar is the RMS of the fit. 
All the data were recorded at $\SI{30}{\milli \kelvin}$ of bath temperature.}
\label{fig:SI_device_extra2}
\end{figure}

In this section we repeated the same magnetic characterization of the Josephson phase battery shown in Fig.~2 and 3 of the main text, performed on a different device to demonstrate the high reproducibility of the effect, apart sample-specific details. Notice that the behaviour of $\varphi_{tot}$ and $\varphi_{int}$ (Fig.~\ref{fig:SI_device_extra1}) is qualitatively similar, but with a smaller total phase shift of $\sim 0.4 \pi$ stemming for a weaker exchange interaction induced by the unpaired-spin. Moreover the angle dependence of $\varphi_{ex}(\theta)$ shown in Fig.~\ref{fig:SI_device_extra2} is in very good agreement with the evolution observed in Fig.~2 and expected from the model presented in Section~\ref{sec:theory}. 

\section{Low magnification SEM image of the device}
\begin{figure*}[h]
\includegraphics[width=\textwidth]{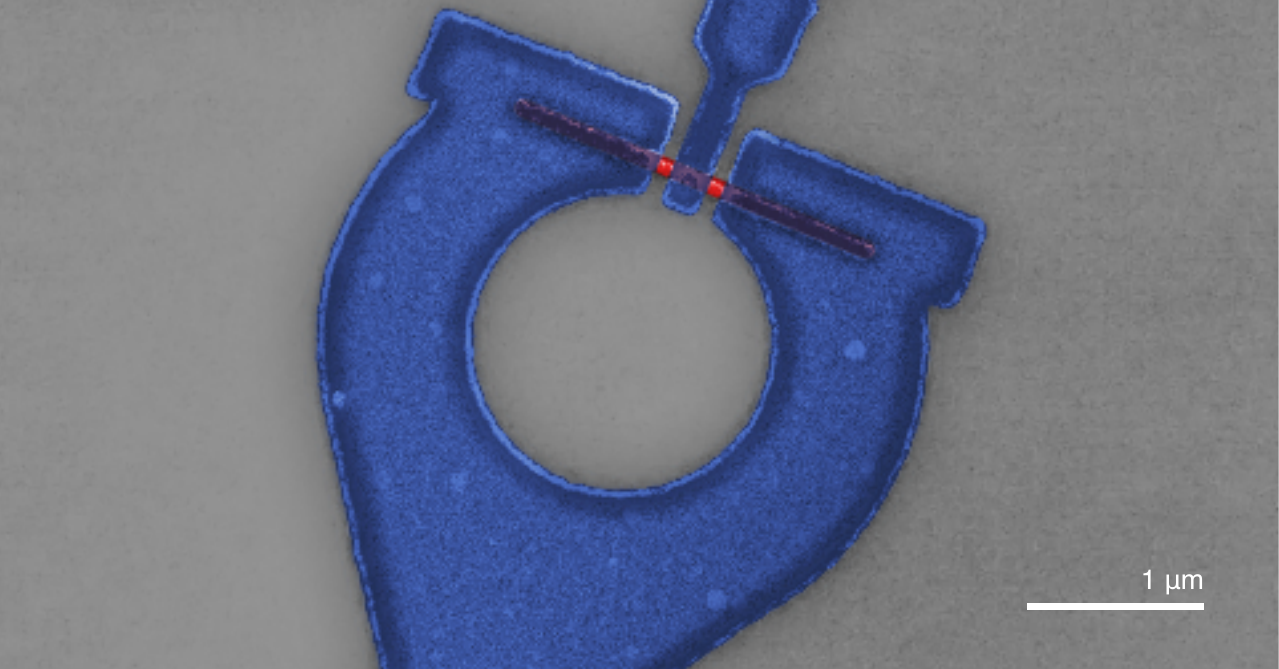}
\caption{\textbf{Low magnification SEM image of the device.} As shown in Fig.~1c of the main text.}
\end{figure*}


\bibliographystyleSI{style}
\bibliographySI{bibliography.bib}

\end{document}